\documentclass{aastex}
\usepackage{emulateapj5}

\newif\ifAMStwofonts
\AMStwofontstrue

\def\kms{km~s$^{-1}$}

\def\ga{\mathrel{\hbox{\rlap{\hbox{\lower4pt\hbox{$\sim$}}}\hbox{$>$}}}}
\def\la{\mathrel{\hbox{\rlap{\hbox{\lower4pt\hbox{$\sim$}}}\hbox{$<$}}}}

\shorttitle{Velocity and density spectra of the SMC}

\shortauthors{S.\ Stanimirovi\'{c} \& A. Lazarian}

\begin{document}

\submitted{Accepted for publication in the Astrophysical Journal Letters}
\title{Velocity and density spectra of the Small Magellanic Cloud}

\author{S.\ Stanimirovi\'{c}}
\affil{Arecibo Observatory, NAIC/Cornell University, HC 3 Box 53995,
Arecibo, Puerto Rico 00612}
\email{sstanimi@naic.edu}
\author{A. Lazarian}
\affil{University of Wisconsin, 475 North Charter Street, 5534 Sterling
Hall, Madison, Wisconsin 53706}

\begin{abstract}

This paper reports results on the statistical analysis of HI turbulence
in the Small Magellanic Cloud (SMC).
We use 21~cm channel maps, obtained with the Australia Telescope
Compact Array and the Parkes telescope, and analyze the spectrum of observed
intensity fluctuations as a function of the velocity slice thickness.
We confirm
predictions by Lazarian \& Pogosyan (2000) on the change of the power law index
and establish the spectra of 3-D density and velocity. The obtained spectral
indices, $-3.3$ and $-3.4$, are slightly
more shallow than the predictions for the Kolmogorov spectrum. This
contrasts to the predictions for the shock-type spectra that are
steeper than the Kolmogorov one. The nature of the energy injection
in the SMC is unclear as no distinct energy injection scales are observed
up to the entire scale of the SMC.
\end{abstract}


\section{Introduction}

Many observations in the past decade have challenged the
traditional picture of the interstellar medium (ISM). Instead of
a two-level hierarchical system,
consisting of clouds uniformly dispersed in the intercloud medium (the so
called `standard cloud' model, Spitzer 1978), the ISM
shows an astonishing inhomogeneity, with many levels of hierarchy.
In order to consider real density functions in physical processes, a better
understanding of the inventory and topology of the ISM is essential, as
well as of the processes responsible for their creation.
Having an extremely gas-rich ISM, dwarf irregular galaxies are
particularly suitable for such studies. Here we discuss
the inventory of the ISM in the Small Magellanic
Cloud (SMC), using the spatial power spectrum, and point to several
processes that may be involved in the sculpturing of its ISM.

The SMC is a nearby\footnote{We assume the SMC to be
at the distance of 60 kpc throughout
this study.}, extremely gas-rich,
dwarf irregular galaxy. Taking
a part in an interacting  system of galaxies (with other members being the
Large Magellanic Cloud, LMC, and our Galaxy), the SMC's morphology,
dynamics and evolution is very complex.
Being one of our closest neighbors, the SMC is attractive for  various
astrophysical aspects. The inventory of the cool
ISM in the SMC was recently studied using the high resolution radio
observations of neutral
hydrogen (HI) \citep{Stanimirovic99}. Also, the dust properties of the SMC
were investigated in detail, and the relationship between the  cool gas and
dust, using both HI and infrared (IR) observations \citep{Stanimirovic00}.

As a dwarf, irregular galaxy, the SMC is different from our own
Galaxy in  many respects: its interstellar radiation field (ISRF)
is  more than four times stronger \citep{Lequeux89}, its heavy
element abundance is almost ten times lower
\citep{Sauvage91}, its dust grains are on average smaller by
$\approx$ 30\%  \citep{Rodrigues97} and significantly hotter
\citep{Stanimirovic00},
and the cool atomic phase of HI is only half as abundant \citep{Dickey99}.
In spite of all the above, the SMC
appears to have similar interstellar turbulence properties,
in having similar power law
index as the Galaxy for the 2-D spatial power spectrum of its HI and dust
column density distributions \citep{Stanimirovic00}.

What do the 2-D spatial power spectra of the intensity fluctuations mean?
Relating the fluctuations of intensity in PPV
(position-position-velocity) data cubes and the underlying 3-D
velocity and density statistics is a problem that has been recently
addressed in \citet{Lazarian99}, where it was shown that changing
the thickness of the velocity slice it is possible to recover both
spectra of turbulent velocity and density.  The SMC provides an ideal testing
ground for this theory and we apply theoretical predictions to the data.
Elsewhere we plan to apply alternative tools for turbulence studies,
e.g. Principal Component Analysis
(PCA) \citep{Heyer97,Brunt01}, the $\Delta$-variance \citep{Stutzki98}
and the spectral correlation function
\citep{Rosolowsky99} (see review by Lazarian 1999) to the SMC data.

The structure of this paper is organized as follows.
In Section~2 we describe briefly the HI observations of the SMC. In
Section~3 we review previous results of the statistical investigation of HI
in the SMC, using the spatial power spectrum, and investigate the influence
of velocity fluctuations on the intensity statistics. Comparison with
earlier work and discussion on the origin of the turbulence in the SMC, are
given in Section~4.

\section{HI Data}
\label{s:hi-data}

The small-scale HI structure in the SMC was observed with the
Australia Telescope Compact Array (ATCA), a radio interferometer, in
a mosaicing mode \citep{Staveleyetal97}. Observations of the same
area were obtained also with the
64-m Parkes telescope, in order to map the distribution of large-scale
features. Both sets of observations were then combined
(Stanimirovic et al. 1999), resulting in the final HI data cube with the
angular resolution of 98 arcsec, the velocity resolution of 1.65 \kms, and
the 1-$\sigma$ brightness-temperature sensitivity of 1.3 K, to the 
continuous range of spatial scales between 30 pc and 4 kpc.
The velocity range covered with these observations is 90
to 215 \kms. For details about the ATCA and Parkes observations, data
processing and data combination (short-spacings correction) see
Staveley-Smith et al. (1997) and Stanimirovic et al. (1999).

\section{Statistical investigation}

The 2-D spatial power spectrum, $P({\bf k})$, of the HI emission
fluctuations ($I$) in
the SMC was first derived in \citet{Stanimirovic99}. This was the first
such study for an entire galaxy.
The power spectrum is defined as the Fourier
transform of the autocorrelation function of the HI emission fluctuations:
\begin{equation}
\label{e:power}
P({\bf k})=\int \int \langle I({\bf x})I({\bf x'}) \rangle e^{-i
{\bf L \cdot k}} d{\bf L}, {\bf L}={\bf x}-{\bf x'}
\end{equation}
with ${\bf k}$ being the spatial frequency, measured in wavelengths,
and ${\bf L}$ being the distance between two points.
To derive $P({\bf k})$ the channel maps (approximately 5 \kms~wide)
were Fourier transformed. The average value of the square of the modulus
of the transform, $\langle
\Re^2 + \Im^2 \rangle$, was then measured in 18 annuli of equal width in
$ \log \sqrt{u^2 + v^2} $ ($u$ and $v$ are the
coordinates in the Fourier plane, being measured in wavelengths) assuming
radial symmetry.
It was shown that the 2-D power spectra were remarkably well fitted
by a power law, $P(k) \propto k^{\gamma}$, over the continuous
range of spatial scales
$\sim$ 30 pc -- 4 kpc, and over the velocity range 110 -- 200 \kms~(several
noisy channels, without significant line emission, were excluded from
the analyses). The
power-law index was remarkably constant over the given velocity
range, and the average slope was estimated with
$\langle \gamma \rangle=-3.04\pm0.02$. No change of the power law slope
was seen on either large or small-scale end of the spectrum.

However, an examination of the
power spectrum of the HI column density distribution, obtained
after integrating along the whole velocity range, showed a change of
the power law slope, with $\gamma$ being equal $-3.31\pm0.01$
\citep{Stanimirovic00}.

The power-law behavior of the spatial power spectrum was seen before
in the case of our Galaxy, for several ISM tracers
\citep{Crovisier83,Green93,Gautier92,Waller96,Schlegel98,Deshpande00b},
and very recently in the case of the LMC \citep{Elmegreen00a}.
The integrated HI column density distribution of the SMC, has a steeper
spatial power spectrum ($\gamma=-3.3\pm0.01$), which is consistent with the
spectrum of the dust column density ($\gamma=-3.1\pm0.2$)
\citep{Stanimirovic00}. In several cases,
a steepening of the slope for higher spatial frequencies was found
\citep{Crovisier83,Waller96,Elmegreen00a}. A change of power-law slope when
examining channel maps and the
integrated intensity along the whole velocity range, noticed in
\citet{Stanimirovic00}, was not found previously.

\subsection{Velocity modification of the HI power spectrum}

The 2-D spatial power spectrum of the intensity (or emissivity)
fluctuations provides important information about the underlying 3-D
statistical properties of the ISM. However, as the spectral-line data cubes
have two spatial, and one velocity axis, both 3-D density and velocity
fluctuations can contribute to the 2-D statistics.
Indeed, due to
the velocity fluctuations, two clumps along the same line of sight at
different distances may appear in the same velocity channel, hence
doubling the measured intensity. In order to relate the 2-D statistics to the
underlying 3-D statistics of the ISM, one must
disentangle density from velocity influence to the power spectrum.

The velocity statistics is probably the most interesting and significant,
partly because it directly represents the dynamics of the media,
and partly because other methods of turbulence study, i.e. scintillation
technique \citep{Spangler99}, fail to deliver this part of information.
The problem of disentangling velocity and density fluctuations
was addressed in \citet{Lazarian99} (see also Lazarian 1999 for a
review of the subject). They found that the relative influence
of velocity and density fluctuations on the intensity fluctuations
 depends on the spectral index of density fluctuations
and on the thickness of channel maps. The latter is easy to understand
qualitatively, as it is clear that line integration should decrease
the influence of velocity fluctuations. The former is less intuitive
but \citet{Lazarian99} showed that if the 3-D density spectrum
is $P_{s} \propto k^{n}$,
two distinct regimes are present when: (a) $n>-3$, and (b) $n<-3$.

Analytical results of  \citet{Lazarian99} study are summarized in
Table~\ref{t:lazarian+pogosyan}. It is easy to see that
in both cases (a) and (b) the power law index
{\it gradually steepens} with the increase of velocity slice
thickness. In the thickest velocity slices the velocity information
is averaged out and it is natural that we get the
density spectral index $n$. The velocity is the most important in
thin slices, and if the 3-D velocity power spectrum is $k^{-3-m}$,
then the index $m$ can be found from thin slices. Note that the
notion of thin and thick slices depends on a turbulence scale under
study and the slice can be thick for small scale turbulent fluctuations
and thin for large scale ones. The formal criterion for the slice to be
thin is that {\it the dispersion of turbulent velocities on the scale studied
should be less than the velocity slice thickness}.  The slice is {\it thick}
otherwise.
When velocity information is averaged out (for a more precise definition
see Lazarian \& Pogosyan 2000) the slice gets {\it very thick} and the velocity
fluctuations get independent of density.

Predictions of \citet{Lazarian99} were successfully tested using numerical
MHD data in \cite{Semadeni00} and thus we can confidently apply  the
theory to observations.

\begin{figure*}
\plotone{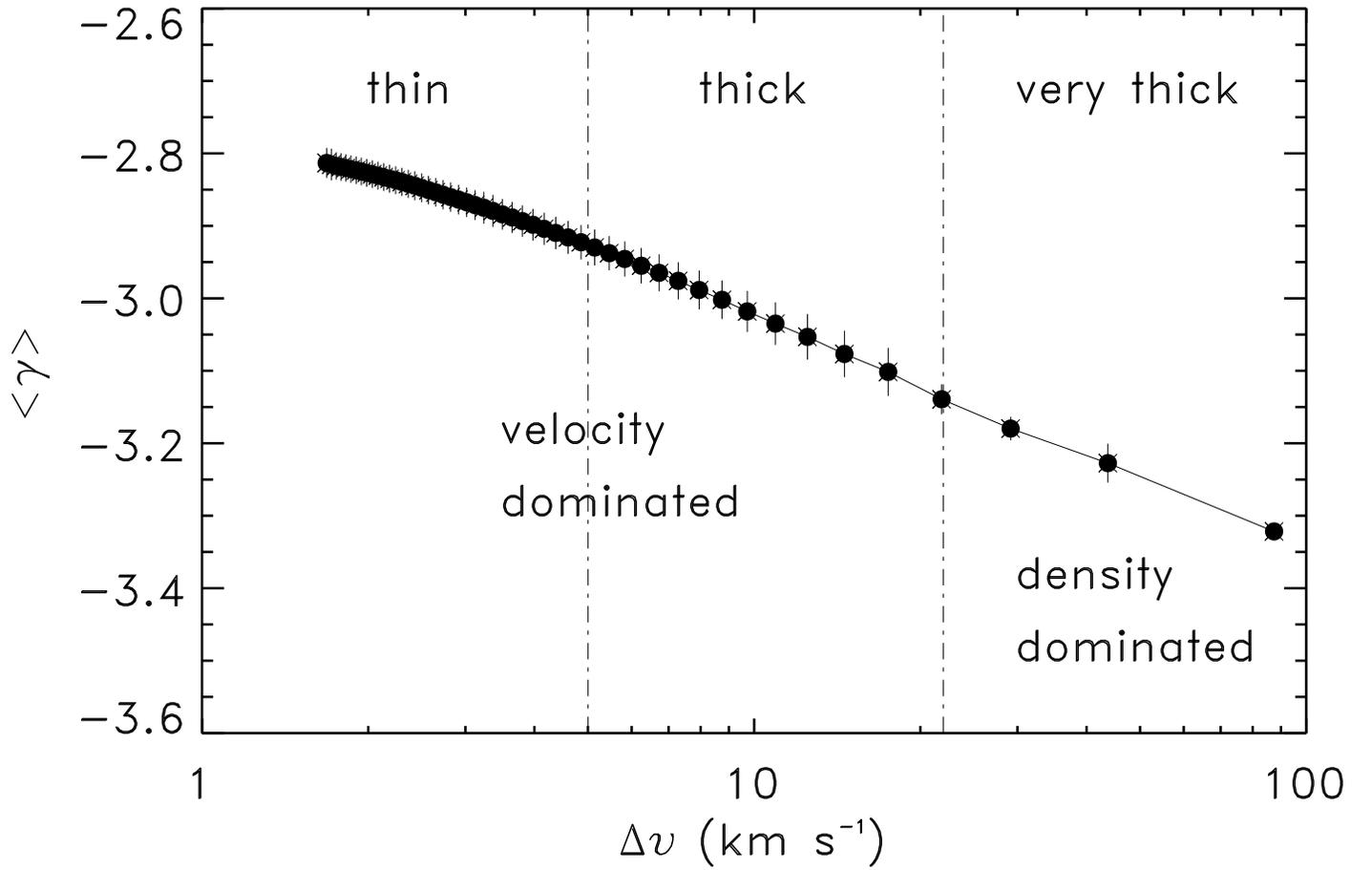}
\caption{\label{f:gamma-variation} The variation of the 2-D HI power spectrum
slope $\langle \gamma \rangle$ with the velocity slice thickness $\Delta
v$. The dot-dashed lines distinguish thin, thick and very thick slice regimes.}
\end{figure*}

To test the predictions by \citet{Lazarian99} in the case of the
SMC, we have determined the 2-D power spectrum slope, $\langle \gamma \rangle$,
while varying the velocity slice thickness from $\sim$ 2 \kms~to $\sim$
100 \kms. A significant variation of $\langle \gamma \rangle$ was found,
shown in Fig.~\ref{f:gamma-variation}, consistent with the predictions ---
$\langle \gamma \rangle$ gradually decreases
with an increase of velocity slice thickness. The thickest velocity slice
gives $n\approx-3.3$, which corresponds to the 3-D density power index
(see Table~1). Using  thin slices, however, we can find the slope of velocity
fluctuations to be $m\approx 0.4$, which means that the 3-D velocity
spectrum index is $-3.4$, which is very close to the 3-D density spectral
index. The transition point between density and velocity dominated
regimes is equal to the velocity dispersion on the scale of the whole SMC
($\sim$ 4 kpc), which is $\sim$ 22 \kms~(Stanimirovic et al., in preparation).

\section{Discussion}

As the spatial power spectrum shows the importance of structure on various
spatial scales, its power-law behavior suggests
the hierarchical structure organization in the ISM, without preferred
spatial scales.
This phenomenon is usually ascribed to the interstellar
turbulence \citep{Scalo87,Elmegreen00}. However, without velocity
information one can always
wonder whether we deal with a static structure or a real turbulence. Indeed, a
distribution of sizes of sand grains on a beach also follows a power-law,
but no one would call this ``turbulence''. The velocity information changes
the picture dramatically. Hence, the extreme importance of the techniques
which relate the observed 2-D power spectrum with the underlying 3-D
statistics of both density and velocity.
Here, we have tested the theoretical predictions for such a technique
\citep{Lazarian99}, and
as a result proved, for the first time, the presence of an active turbulence
in the SMC.

In view of theoretical results in \citet{Lazarian99} it is now appropriate
to reanalyze all the earlier data. These data
were obtained without much concern about the thickness of velocity slices.
Therefore the observed variations of the power index can be due to
transitions from ``thin'' to ``thick'' and to ``very thick'' slices.
In the case of Green's (1993) data additional complications are
related to a divergent line of sight geometry. A more detailed discussion
of the available data will be given elsewhere.
We note that although the description in terms of power spectra is common in
hydrodynamics and the MHD theory, it has certain limitations, as discussed in
Lazarian (1999). For example, the power spectrum analysis does not include
information about the phase distribution, dealing only with the modulus
of the Fourier transform, nor it contains information about the
structure connectivity (Scalo 1987). Other methods hence,
should be used as complementary statistical descriptors. Power
spectrum as it is can provide us with an important
insight of what kind of turbulence
we deal with, e.g. distinguish the turbulence originating from
shock waves from the hydrodynamic turbulence.

The attempts to test Lazarian \& Pogosyan (2000) theory were made
recently in Elmegreen et al. (2000) using the HI observations of the LMC.
In agreement
with theoretical predictions, the steepening of spectrum was observed
for high spatial frequencies. The puzzling thing discovered by Elmegreen
et al. (2000) was the flattening of the spectra for velocity-integrated
intensity, which was interpreted as an effect of the finite LMC
disk thickness. This is an interesting explanation which entails that
the LMC spectrum at the scales larger than 100~pc becomes essentially
two dimensional.
Our study has not noticed a systematic change of the velocity
integrated power spectrum at large spatial scales. This may reflect
the fact that the SMC, unlike the LMC, is essentially a 3-D entity.

Another approach in relating the 2-D with the 3-D statistics in the case of
the SMC was presented in Goldman (2000).
There it is assumed that the density fluctuations are a ``passive
scalar'', being driven by the velocity fluctuations, and hence having the
same power spectrum. If we accept that the intensity fluctuations are due
to the density fluctuations, then the corresponding spectral
index  $q$ of intensity fluctuations in a 2-D slice
can then be related to the 3-D density spectrum index $n$ as
\begin{equation}
q=n+1~~~,
\label{q}
\end{equation}
which for the SMC data produces $n\approx -4$\footnote{This would
correspond to the spectrum of shock waves. To avoid possible confusion
we point out that we talk about power spectrum which differs from
the energy spectrum by $k^2$. The Kolmogorov turbulence corresponds
to the power spectrum of $-11/3$.}.
However, we note that the data used in Goldman (2000) are not in the real
space ($xyz$) for which his treatment would be
correct, but are in the velocity space ($xyv$). In this situation
the  \citet{Lazarian99} treatment is appropriate and it provides a different
result, namely, velocity index $\approx -3.4$ and density index
$\approx -3.3$. Eq.~(\ref{q}) also predicts that the difference in the
power slope between thin and thick slices is equal to $1$. This is
inconsistent with Fig.~1.
We also note that for the Kolmogorov spectrum the predictions
in \citet{Lazarian99} (see Table~1) coincide with $n$ calculated using
Eq. (\ref{q}), but this correspondence is accidental.

An interesting application of the power spectrum of HI opacity fluctuations
was made by \citet{Deshpande00a},
in order to explain the long-standing puzzle of the tiny-scale
structure in HI \citep{Heiles97}.
Assuming a single power
spectrum of the opacity fluctuations, with a slope of 2.75 over the range of
$\sim$ 0.02 pc to $\sim$ 4 pc, \citet{Deshpande00a} obtained opacities
consistent with the observations of small-scale HI structure in
\citet{Deshpande00b}. This is very
encouraging and requires re-interpretation of previous observations of the
small-scale structure in a similar way. A preliminary investigation
of the small-scale structure found so far by \citet{Heiles00} suggested though
a more complex structure function, with
a significant change of slope for scales smaller than 0.01 pc.

What can drive the turbulence in the SMC? The natural assumption
would be that it is due to the stirring of the ISM produced by
a large number of expanding shells found in the SMC.
The shell sizes range from $\sim$ 30 pc to $\sim$ 2 kpc.
Hence, one scenario could be that the
largest shells drive the turbulent cascade down to smallest observed
scales. However, processes like shell fragmentation and/or shell
propagation from the smaller scales (bottom-up scheme, see Scalo 1987), may play
significant role too.
The main problem in pinning down the exact mechanisms is that,
so far, we have not observed changes in the power spectrum,
at any scale up to the entire size of the SMC, which would be indicative
of energy injection. An alternative explanation was suggested in
\citet{Goldman00}, whereby the large scale turbulence is induced by
instabilities in the large-scale flows during the last SMC--LMC encounter.
Future comparison with simulations of different types of turbulent
cascades are essential to resolve this question.

\section{Conclusions}
We have successfully tested predictions of the \citet{Lazarian99}
study on the change of slope of the intensity fluctuation spectrum
with the velocity slice thickness. The SMC spectrum appears
due to active turbulent motions rather than just a static
hierarchical structure. We found the index for the 3-D velocity spectrum
to be $-3.4$, while for the density spectrum to be $-3.3$.

\begin{table*}
\caption{\label{t:lazarian+pogosyan}A summary of analytical results for the
2-D power spectrum of intensity fluctuations, $P(k)$, derived
in Lazarian \& Pogosyan (2000) for two distinct regimes of the 3-D density
fluctuations. The index of the 3-D power spectrum of velocity fluctuations 
is $-3-m$.}
\begin{tabular}{lcc}
\noalign{\smallskip} \hline \hline \noalign{\smallskip}
Slice & Shallow 3-D density & Steep 3-D density\\
thickness & $P_{s} \propto k^{n}$, $n>-3$&$P_{s} \propto k^{n}$, $n<-3$\\
\noalign{\smallskip} \hline \noalign{\smallskip}
2-D intensity spectrum for thin\tablenotemark{a}~~slice &
$\propto k^{n+m/2}$    & $\propto
k^{-3+m/2}$   \\
2-D intensity spectrum for thick\tablenotemark{b}~~slice & $\propto k^{n}$
& $\propto k^{-3-m/2}$  \\
2-D intensity spectrum for very thick\tablenotemark{c}~~slice & $\propto k^{n}$ & $\propto k^{n}$  \\
\noalign{\smallskip} \hline \noalign{\smallskip}
\end{tabular}
\tablenotetext{a}{channel width $<$ velocity dispersion at the scale under
study}
\tablenotetext{b}{channel width $>$ velocity dispersion at the scale under
study}
\tablenotetext{c}{substantial part of the velocity profile is integrated over}
\end{table*}

\label{lastpage}
\end{document}